\documentclass[10pt]{llncs}
\usepackage{times}

\usepackage{algpseudocode}
\usepackage{algorithm}

\usepackage{xyling}
\usepackage{tikz}  
 \usetikzlibrary{fit}
\usetikzlibrary{trees}
\usetikzlibrary{arrows}
\usetikzlibrary{decorations.markings}

\usepackage{etex}

\usepackage{amsmath,amssymb}
\usepackage{mathabx}
\usepackage{subfig}
\usepackage{booktabs}
\usepackage[normalem]{ulem}
\usepackage{colortbl}
\usepackage{listings}
\usepackage{chngcntr}
\usepackage{url}
\usepackage{paralist}
\usepackage{fancybox}
\usepackage{float}
\algrenewcommand\algorithmicindent{0.4em}%
\usepackage{xspace}
\def\fedra{\textsc{Fedra}\xspace}

\newcolumntype{C}[1]{>{\centering\let\newline\\\arraybackslash\hspace{0pt}}m{#1}}
\usepackage[utf8]{inputenc}

\begin{document}

\title{Efficient Query Processing  for SPARQL Federations with Replicated Fragments}
\author{Gabriela Montoya\inst{1} \and Hala Skaf-Molli \inst{1} \and Pascal Molli \inst{1}  \and Maria-Esther Vidal \inst{2}}
\institute{
LINA-- Nantes University, France \email{\{first.last\}@univ-nantes.fr}
\and  Universidad Sim\'on Bol\'{\i}var, Venezuela \email{mvidal@ldc.usb.ve}
}

\maketitle

 \begin{abstract}
   Low reliability and availability of public SPARQL endpoints prevent
   real-world applications from exploiting all the potential of these
   querying infrastructures. Fragmenting data on servers can improve
   data availability but degrades performance. Replicating fragments
   can offer new tradeoff between performance and availability. We
   propose \fedra, a framework for querying Linked Data that takes
   advantage of client-side data replication, and performs a source
   selection algorithm that aims to reduce the number of selected
   public SPARQL endpoints, execution time, and intermediate
   results. \fedra has been implemented on the state-of-the-art query
   engines ANAPSID and FedX, and empirically evaluated on a variety of
   real-world datasets. 
 \end{abstract}


\section{Introduction}

Linked Data~\cite{DBLP:journals/ijswis/BizerHB09} provides millions of
triples for data consumers, however, recent studies suggest that
data availability is currently the main bottleneck to the success of the
Semantic Web as a viable
technology~\cite{DBLP:conf/semweb/ArandaHUV13,verborgh_iswc_2014}.
Particularly, it has been reported by Buil-Aranda et
el.~\cite{DBLP:conf/semweb/ArandaHUV13} that only one third of the 427
public SPARQL endpoints have an availability rate equal or greater
than 99\%; representing this limitation the major obstacle to
developing Web real-world applications that access Linked Data by
using these  infrastructures.
Recently, the Linked Data Fragments (LDF) approach~\cite{verborgh_ldow_2014,verborgh_iswc_2014}
has addressed availability issues by delegating query processing to
clients, and by transforming public endpoints into simple HTTP-based
triple-pattern fragments providers that can be easily cached by
clients. This tradeoff effectively improves data availability, but it
can significantly degrade performance~\cite{verborgh_ldow_2014}. 
However, we speculate that fragments caching approaches like LDF can have  
two important consequences for consuming Linked Data:
\begin{inparaenum}[(i)]
\item Each client is able to process SPARQL queries on replicated
  fragments cached from servers. Consequently, if clients are
  ready to cooperate, the cost of executing SPARQL queries can be
  significantly reduced, and a new compromise between availability and
  performance can be achieved.
\item Potentially, clients could cache  triple-pattern fragments from different
  data providers creating new localities for federated
  queries. Therefore, some joins can  be now executed on one
  machine without contacting the public endpoints.
\end{inparaenum}
This vision is also clearly proposed by Iba\~nez~\cite{colgraph}, where
triple-pattern fragments can be replicated from SPARQL public
endpoints, modified locally, and made available through consumer data
endpoints.
Approaches in~\cite{colgraph,verborgh_ldow_2014}  demonstrate how SPARQL processing resources and
simple triple-based fragments can be obtained from data
consumers.  We believe this represents a new opportunity for federated query
processing engines to improve SPARQL query processing performance 
by taking advantage of opportunistic replication and SPARQL processing
offered by data consumers. 

%
%
\begin{table}[t]
\centering
\begin{tabular}{|c|r|r|r|r|}
\hline
\#DBpedia & \multicolumn{2}{c|}{Execution Time (ms)} & \multicolumn{2}{c|}{\# Results} \\
\cline{2-5}
 Replicas  & FedX& ANAPSID & FedX & ANAPSID\\
\hline
1 & 1,392 & 22,972 & 8,921 & 8,921\\
\hline
2 & 215,907 & $\geq$ 1,800,000 & 418 & 8,921\\
\hline
\end{tabular}
\caption{Execution time and results for the same query over one and two
    replicas of DBpedia for FedX and ANAPSID}
 \label{fig:2dbpedia}
 \vspace{-0.75cm}
\end{table}

However, current SPARQL federated query
engines~\cite{DBLP:conf/semweb/AcostaVLCR11,DBLP:conf/semweb/BascaB10,DBLP:conf/semweb/GorlitzS11,DBLP:conf/esws/QuilitzL08,DBLP:conf/semweb/SchwarteHHSS11} may exhibit poor
performance in presence of replication. As presented in Figure~\ref{fig:2dbpedia}, we duplicated DBpedia and executed 
a three triple pattern query against one instance and next two instances of DBpedia. We can observe that the performance in terms of execution time
and number of results is seriously degraded.
This problem has been partially addressed by recent duplicate-aware source selection 
strategies~\cite{DBLP:conf/sigmod/HoseS12,DBLP:conf/semweb/SaleemNPDH13}.
The proposed solutions rely
on summary of datasets to detect overlapping and do not
consider fragments~\cite{colgraph,verborgh_iswc_2014,verborgh_ldow_2014}. With
fragments, replication is defined declaratively and does not need to be detected.

In this paper, we propose \fedra, a source selection strategy 
that exploits fragment definition to select 
non-redundant data sources.  In contrast to ~\cite{DBLP:conf/sigmod/HoseS12,DBLP:conf/semweb/SaleemNPDH13}, \fedra does not
require information about the content of the data sources to detect
overlapping.  \fedra just relies on knowledge about the endpoint
replicated fragments to reduce the number of endpoints to be 
contacted, and delegate join execution to endpoints.
 \fedra  implements  a set covering heuristic ~\cite{DBLP:books/daglib/0017733} to minimize
the number of sources to be contacted. The implemented source selection approach ensures that triple patterns in the same basic 
graph pattern are assigned to the same endpoints, and consequently, it reduces
the size of intermediate results.
We extend the state-of-the art federated query engines 
FedX~\cite{DBLP:conf/semweb/SchwarteHHSS11} and
ANAPSID~\cite{DBLP:conf/semweb/AcostaVLCR11} with \fedra, and compare
these extensions with the original engines. 
We empirically study these engines and the results suggest that 
\fedra efficiently reduces the number of public, replicated 
endpoints, and intermediate results.
The paper is organized as follows: 
Section~\ref{sec:relatedwork} presents related works. 
Section~\ref{sec:approach} describes \fedra and the source selection algorithm.
Section~\ref{sec:experiments} reports our experimental results. 
Finally, conclusions and future works are outlined in Section~\ref{sec:conclusion}.

\section{Related Work}
\label{sec:relatedwork}

In distributed databases, data fragmentation and
replication improve data availability and
query performance~\cite{ozsu2011principles}. 
Linked Data~\cite{DBLP:journals/ijswis/BizerHB09} is intrinsically a
federation of autonomous participants where federated queries are
unknown from a single participant, and a tight coordination of data
providers is difficult to achieve. This makes data
fragmentation~\cite{ozsu2011principles} and distributed query
processing~\cite{DBLP:journals/csur/Kossmann00} of distributed
databases not a viable
solution~\cite{DBLP:journals/www/UmbrichHKHP11} for Linked Data.
 
Recently, the Linked Data fragments approach
(LDF)~\cite{verborgh_ldow_2014,verborgh_iswc_2014} proposes to improve
Linked Data availability by moving query execution load from servers to
clients. A client is able to execute locally a restricted SPARQL query
by  downloading  fragments required to execute the query from an
LDF server through a simple HTTP request. This strategy allows clients to
cache fragments locally and decreases the load on the LDF
server. \emph{A Linked Data Fragment of a Linked Data dataset is a
  resource consisting of the dataset triples that match a
  specific selector}. A triple pattern fragment is a special kind of
fragments where the selector is a triple pattern; a triple pattern
fragment minimizes the effort of the server to produce the
fragments. LDF  chose a clear tradeoff by shifting query processing
to clients, at the cost of slower query
execution~\cite{verborgh_iswc_2014}. On the other hand,
 LDF could create many data consumers resources that hold
replicated fragments  in their cache and is able to process SPARQL
queries. This opens the opportunity to use these new resources to
process SPARQL federated queries. \fedra aims to improve source
selection algorithm of federated query engine to consider these new
endpoints, and decreases the load on public endpoints.

Col-graph~\cite{colgraph} enables data consumers to materialize triple
pattern fragments and to expose them through SPARQL endpoints to
improve data quality.  A data consumer can update her local fragments
and share updates with data providers and consumers. Col-graph proposes a coordination
free protocol to maintain the consistency of replicated
fragments. Compared to LDF, Col-graph clearly creates SPARQL endpoints
available for other data consumers, and allows federated query engines to
use local fragments. As for LDF, \fedra can take advantage of these
data consumer resources. 

Recently, HiBISCuS~\cite{DBLP:conf/esws/SaleemN14} a source selection
approach has been proposed to reduce the number of selected sources.
The reduction is achieved by annotating sources with the URIs
authorities they contain, and pruning sources that cannot have triples
that match any of the query triple patterns.  HiBISCuS differs from our aim 
of both selecting
sources that are required to the answer, and avoiding the selection of
sources that only provide redundant replicated
fragments. While not directly related to replication, HiBISCuS index
could be used in conjunction with \fedra to perform join-aware source
selection in presence of replicated fragment.


Existing federated query engines
~\cite{DBLP:conf/semweb/AcostaVLCR11,DBLP:conf/semweb/BascaB10,DBLP:conf/semweb/GorlitzS11,DBLP:conf/esws/QuilitzL08,DBLP:conf/semweb/SchwarteHHSS11} are not able
to take advantage of replicated fragments, and data overlapping can
seriously degrade their
performance as reported in Figure~\ref{fig:2dbpedia} and
shown in~\cite{Fedra,DBLP:conf/semweb/SaleemNPDH13}. We integrated \fedra
within FedX and ANAPSID to make existing engines aware of replicated
fragments. With \fedra, replications as in
Figure~\ref{fig:2dbpedia} will be detected, and performance will
remain stable.

Recently, QBB\cite{DBLP:conf/sigmod/HoseS12} and
DAW\cite{DBLP:conf/semweb/SaleemNPDH13} propose duplicate-aware
strategies for selecting sources for federated query engines. Both
approaches use sketches to estimate the overlapping among sources. 
DAW uses a combination of Min-Wise Independent Permutations 
(MIPs)~\cite{DBLP:journals/jcss/BroderCFM00}, and triple selectivity
information to estimate the overlap between the results of different
sources.  Based on how many new query results are expected to be
found, sources below predefined benefits are discarded and not
queried. 

Compared to DAW, \fedra does not require to compute data summaries 
because \fedra relies on fragment definitions to deduce
containments. Computing containments based on fragment descriptions is less
expensive than computing data summaries, moreover data updates
are more frequent than fragment description updates.
\fedra also try to remove public endpoints and minimize
the number of endpoints to execute a query. Consequently, if DAW and
\fedra could  find the same number of sources to execute a query, 
\fedra minimizes the number of
public endpoints to be contacted. Additionally, \fedra source selection
considers the query basic graph patterns to delegate join execution 
to the endpoints and reduce intermediate
results size. This key feature cannot be achieved by DAW as it performs source
selection only at the triple pattern level.
\tikzset{
  big arrow/.style={
    decoration={markings,mark=at position 1 with {\arrow[scale=2,#1]{>}}},
    postaction={decorate},
    shorten >=0.4pt},
  big arrow/.default=blue}

\section{Fedra Approach}
\label{sec:approach}

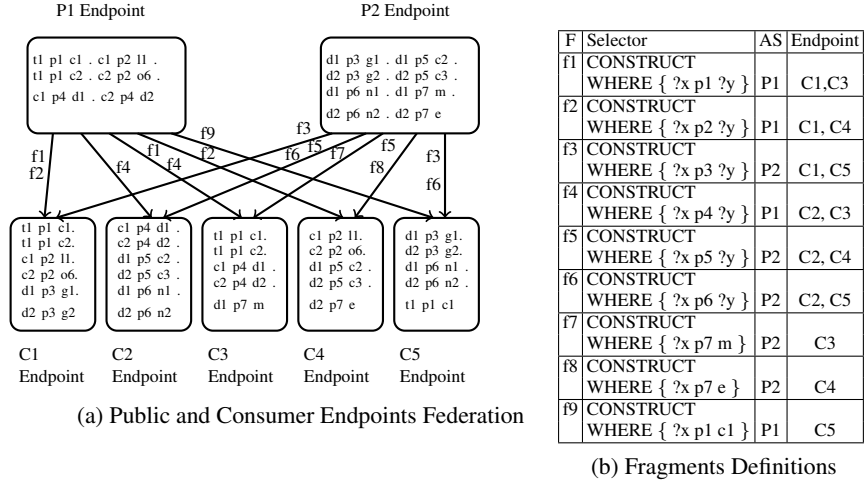
\begin{figure}[t]
\subfloat[][Public and Consumer Endpoints Federation]{%
\label{fig:replica-1}
\begin{tikzpicture}[thick,scale=0.75, every node/.style={scale=0.75}]
 \draw [rounded corners] (0.3,2.5) rectangle (3.1,4.2);    
 \draw [rounded corners] (5.5,2.5) rectangle (8.3,4.2);   
 \draw [rounded corners] (0,-1) rectangle (1.5,1);   
 \draw [rounded corners] (1.7,-1) rectangle (3.2,1);   
 \draw [rounded corners] (3.4,-1) rectangle (4.9,1);   
 \draw [rounded corners] (5.1,-1) rectangle (6.6,1);   
 \draw [rounded corners] (6.8,-1) rectangle (8.3,1);   
 \draw [thick, ->] (0.75,2.5) -- (0.6,1);
 \draw [thick, ->] (1.25,2.5) -- (2.5,1);
 \draw [thick, ->] (1.75,2.5) -- (4.1,1);
 \draw [thick, ->] (2.25,2.5) -- (5.9,1);
 \draw [thick, ->] (2.85,2.5) -- (7.5,1);
 \draw [thick, ->] (5.7,2.5) -- (0.8,1);
 \draw [thick, ->] (6.3,2.5) -- (2.7,1);
 \draw [thick, ->] (6.6,2.5) -- (4.3,1);
 \draw [thick, ->] (7.2,2.5) -- (6.1,1);
 \draw [thick, ->] (7.7,2.5) -- (7.7,1);
 \node at (1.6,4.65) {P1 Endpoint};
 \node at (7,4.65) {P2 Endpoint};
 \node at (0.5,2.1) {\small f1};
 \node at (0.46,1.8) {\small f2};
 \node at (2,1.9) {\small f4};
 \node at (2.55,2.2) {\small f1};
 \node at (2.9,1.95) {\small f4};
 \node at (3.5,2.15) {\small f2};
 \node at (3.5,2.5) {\small f9};
 \node at (5.2,2.6) {\small f3};
 \node at (5.4,2.27) {\small f5};
 \node at (5,2.13) {\small f6};
 \node at (5.8,2.15) {\small f7};
 \node at (6.7,2.3) {\small f5};
 \node at (6.5,1.9) {\small f8};
 \node at (7.5,2.1) {\small f3};
 \node at (7.5,1.6) {\small f6};
 \node[text width=1cm, text height=1cm] at (0.65,-1.25) {C1 \\ Endpoint};
 \node[text width=1cm, text height=1cm] at (2.3,-1.25) {C2 \\ Endpoint};
 \node[text width=1cm, text height=1cm] at (4,-1.25) {C3 \\ Endpoint};
 \node[text width=1cm, text height=1cm] at (5.7,-1.25) {C4 \\ Endpoint};
 \node[text width=1cm, text height=1cm] at (7.4,-1.25) {C5 \\ Endpoint};
 \node[text width=3.2cm, text height=1.75cm] at (2,4.25) {\scriptsize t1 p1 c1 . c1 p2 l1 . \\ t1 p1 c2 . c2 p2 o6 .\\ c1 p4 d1 . c2 p4 d2  };
 \node[text width=3.2cm, text height=1.75cm] at (7.2,4.1) {\scriptsize d1 p3 g1 . d1 p5 c2 .\\ d2 p3 g2 . d2 p5 c3 .\\ d1 p6 n1 . d1 p7 m .\\ d2 p6 n2 . d2 p7 e };
 \node[text width=3.2cm, text height=1.75cm] at (1.8,0.85) {\scriptsize t1 p1 c1.\\ t1 p1 c2.\\ c1 p2 l1.\\ c2 p2 o6.\\ d1 p3 g1.\\ d2 p3 g2  };
 \node[text width=3.2cm, text height=1.75cm] at (3.5,0.85) {\scriptsize c1 p4 d1 .\\ c2 p4 d2 .\\ d1 p5 c2 .\\ d2 p5 c3 .\\ d1 p6 n1 .\\ d2 p6 n2};
 \node[text width=3.2cm, text height=1.75cm] at (5.2,0.85) {\scriptsize t1 p1 c1.\\ t1 p1 c2.\\ c1 p4 d1 .\\ c2 p4 d2 .\\ d1 p7 m };
 \node[text width=3.2cm, text height=1.75cm] at (6.9,0.85) {\scriptsize c1 p2 l1.\\ c2 p2 o6.\\ d1 p5 c2 .\\ d2 p5 c3 .\\ d2 p7 e };
 \node[text width=3.2cm, text height=1.75cm] at (8.6,0.85) {\scriptsize d1 p3 g1.\\ d2 p3 g2.\\ d1 p6 n1 .\\ d2 p6 n2 .\\ t1 p1 c1 };
\end{tikzpicture}
}\hspace*{-0.75cm}
\subfloat[][Fragments Definitions]{%
\raisebox{20mm}{
\label{tab:replica-2}%
\begin{scriptsize}
\begin{tabular}{|c|l|c|c|} \hline
F & Selector & AS  & Endpoint\\ \hline
f1 & CONSTRUCT  & & \\
   & WHERE \{ ?x p1 ?y \} & P1 &C1,C3 \\  \hline
f2 & CONSTRUCT & & \\
   & WHERE \{ ?x p2 ?y \} & P1 & C1, C4\\ \hline
f3 & CONSTRUCT & & \\
   & WHERE \{ ?x p3 ?y \} & P2 & C1, C5 \\ \hline
f4 & CONSTRUCT & & \\
   & WHERE \{ ?x p4 ?y \} & P1 & C2, C3\\ \hline
f5 & CONSTRUCT & & \\
   & WHERE \{ ?x p5 ?y \} & P2 & C2, C4\\ \hline
f6 & CONSTRUCT & & \\
   & WHERE \{ ?x p6 ?y \} & P2 & C2, C5\\ \hline
f7 & CONSTRUCT & & \\
   & WHERE \{ ?x p7 m \} & P2 & C3\\ \hline
f8 & CONSTRUCT & & \\
   & WHERE \{ ?x p7 e \} & P2 & C4\\ \hline
f9 & CONSTRUCT & & \\
   & WHERE \{ ?x p1 c1 \} & P1 & C5\\ \hline
\end{tabular}
\end{scriptsize}
}
}%
\caption[A federation  composed of two public SPARQL endpoints and
   five consumer endpoints]{ figure~\subref{fig:replica-1} describes
     how fragments of public endpoints are
   replicated at consumer endpoints. Table \subref{tab:replica-2}
   describes the selector of each fragment, "authoritative'' source (AS) and Endpoint where the fragment is available}
\label{fig:replicas}
\vspace{-0.5cm}
\end{figure}


\paragraph{Fragments and Endpoints Descriptions}

To define a fragment, we will use the Linked Data Fragment definition
given by Verborgh et al.~\cite{verborgh_iswc_2014}.  Let $ \mathcal{U} , \mathcal{L}$,
and $\mathcal{V}$ denote the set of all URIS, literals
and variables, respectively.  $\mathcal{T}^*= \mathcal{U} \times \mathcal{U} \times
(\mathcal{U} \cup \mathcal{L)}$ is a finite set of blank-node-free RDF
triples.  Every dataset $G$ published via some kind of fragments on
the Web is a finite set of blank-node free triples; \emph{i.e.}, $G
\subseteq \mathcal{T}^*$.  Any tuple $tp \in (\mathcal{U} \cup
\mathcal{V}) \times (\mathcal{U} \cup \mathcal{V} ) \times
(\mathcal{U} \cup \mathcal{V} \cup \mathcal{L}$) is a triple pattern.
\begin{definition}[Fragment~\cite{verborgh_iswc_2014}]
 A Linked Data Fragment (LDF)  of G is a tuple $f = \langle u, s, \Gamma , M,C \rangle$ with the following five elements:
\begin{inparaenum}[\itshape i\upshape)]
\item  u is a URI (which is the ``authoritative'' source from which f can be retrieved);
\item s is a selector;
\item $\Gamma$  is a set consisting of all subsets of G that match selector s, that is, for every
$G'\subseteq G$ it holds that  $G'  \in \Gamma$  if and only if  $G' \in  dom(s)$ and  $s(G') = true$;
\item  M is a finite set of (additional) RDF triples, including triples that represent
metadata for f; and
\item C is a finite set of controls.
\end{inparaenum}
\end{definition}
%
We restrict fragments to triple pattern fragments as in~\cite{colgraph,verborgh_ldow_2014}.
Hereafter, we consider that fragments are read-only and data cannot be updated;  the
fragment synchronization problem is studied in~\cite{colgraph}.\\
Consider the federation in Figure~\ref{fig:replica-1}, where five data
consumer  endpoints (\textit{C1}-\textit{C5})  include fragments
(\textit{f1}-\textit{f9}) from public endpoints \textit{P1} and
\textit{P2}.  Table~\ref{tab:replica-2} shows the SPARQL CONSTRUCT query used as
a selector for a fragment.  Fragments \textit{f1}, \textit{f2}, \textit{f4}, and \textit{f9} have as
``authoritative'' source \textit{P1}, and fragments \textit{f3} and \textit{f5-f8} have as
``authoritative'' source \textit{P2}. The last column presents the consumer endpoints
where  fragments are available.
To participate in a \fedra federation, data consumers annotate 
each fragment exposed through their endpoints with the fragment selector 
$s$  and the public endpoint that provides the data $u$.
The vocabulary term \textit{sd:endpoint} refers to 
the SPARQL endpoint that publishes the fragment.
The vocabulary term \textit{dcterms:hasPart} introduces a fragment description, the
vocabulary term \textit{dc:description} refers to the SPARQL CONSTRUCT query $s$, and 
the vocabulary term \textit{dcterms:source} specifies $u$, the fragment 
``authoritative'' source URI.
Listing~\ref{lst:endpointDescription} shows the description of
the  endpoint \textit{C1} of Figure~\ref{fig:replica-1}.

\begin{figure}[t]
\noindent\begin{minipage}{.65\textwidth}
\begin{lstlisting}[basicstyle=\tiny\sffamily,language=sparql,caption={C1 Endpoint Description},numbers=none,frame=none,columns=fixed,label=lst:endpointDescription,extendedchars=true,breaklines=true,showstringspaces=false]
@prefix sd:<http://www.w3.org/ns/sparql-service-description#>.
@prefix dc: <http://purl.org/dc/elements/1.1/> .
@prefix dcterms: <http://purl.org/dc/terms/> .
[] <http://www.w3.org/1999/02/22-rdf-syntax-ns#type> sd:Service ;
 sd:endpoint <http://consumer1/sparql>;
 dcterms:hasPart [
  dc:description "Construct where{ ?x p1 ?y }";
  dcterms:source <http://publicEndpoint1/sparql>; ] ;
 dcterms:hasPart [
  dc:description "Construct where{ ?x p2 ?y }";
  dcterms:source <http://publicEndpoint1/sparql>; ] ;
 dcterms:hasPart [
  dc:description "Construct where{ ?x p3 ?y }";
  dcterms:source <http://publicEndpoint2/sparql>; ] ;
\end{lstlisting}
\end{minipage}\hfill
\begin{minipage}{.35\textwidth}
\begin{lstlisting}[basicstyle=\scriptsize\sffamily,language=sparql,breaklines=true,label=lst:queries,caption={Queries Q1 and Q2}]
Q1: CONSTRUCT 
    where { ?x1 p1 ?x2 }
    
Q2: CONSTRUCT 
    where { ?x1 p4 ?x2 . 
            ?x1 p7 ?x3 }
\end{lstlisting}
\end{minipage}
\vspace{-0.75cm}
\end{figure}
For source selection with replicated fragments, we need to define when a fragment  is relevant for answering a query.
A fragment  is relevant for answering a query,  if it is relevant for at least one  triple pattern of the query.
\begin{definition}[Fragment relevance]
A fragment $f = \langle u, s, \Gamma , M,C \rangle$ is relevant for a triple pattern $tp$, 
if the triple pattern evaluated over $\Gamma$, $\ldbrack tp \rdbrack_{\Gamma}$~\cite{DBLP:journals/tods/PerezAG09}, is
not empty.
\label{def:fragmentrelevance}
\end{definition}


Consider queries Q1 and Q2 (cf. Listing~\ref{lst:queries}), and the federation 
of Figure~\ref{fig:replicas}. Fragments \textit{f1} and \textit{f9} are relevant for query \textit{Q1}, 
while fragments \textit{f4}, \textit{f7}, and \textit{f8} are relevant for query \textit{Q2}.
We can define two types of containments: containment between SPARQL
endpoints and containment between fragments.

\begin{definition}[Endpoint  Containment]
Let e1 and e2 be the URI of  two SPARQL endpoints that respectively expose  fragments 
$f1=  \langle u1, s, \Gamma1 , M1,C1 \rangle$ and $f2 = \langle u2$ , $s, \Gamma2 , M2,C2 \rangle$  
such that 
f1 and f2 have the same selector $s$ and  the same ``authoritative'' source   ($u2$ = $u1$). 
Then all triples in f2 are contained in f1 and vice versa, i.e., $\Gamma2 \subseteq \Gamma1 \land \Gamma1 \subseteq \Gamma2$. And we use the notation e1 $\subseteq_s$ e2, and e2 $\subseteq_s$ e1 to represent the endpoint containment relationship.
\label{def:containmenti}
\end{definition}

\textit{f1} triples in consumers \textit{C1} and \textit{C3} from Figure~\ref{fig:replicas} are the same (from Definition \ref{def:containmenti}):
$C1 \subseteq_s C3$ and $C3 \subseteq_s C1$ 
 where $s$ is $f1$ selector.
Endpoint containment can be used to reduce the number of endpoints to
contact to answer a query.  \textit{f1} and \textit{f9} are relevant for
query \textit{Q1}, from the endpoint containments \textit{f1} data 
is also available through \textit{P1}, \textit{C1}, and \textit{C3} and \textit{f9} data is also available
through \textit{P1} and \textit{C5}.  Therefore, only one endpoint per
fragment needs to be selected to answer the query.  A good choice
could be \textit{C1}, \textit{C5} or \textit{C3}, \textit{C5};  this will reduce the load of the public endpoint
\textit{P1} and will improve \textit{P1} availability.  By contacting only \textit{C1}, \textit{C5} or \textit{C3},
\textit{C5}, complete query answers are obtained because the triple
pattern fragment defines a copy of the data source using the fragment
selector.
Another type of containment that allows to reduce the number of
sources to be contacted is defined based on the fragment selector.
\begin{definition}[Fragment Containment]
Let $f1=  \langle u, s1, \Gamma1 , M1,C1 \rangle$ and $f2 = \langle u, s2, \Gamma2 , M2,C2 \rangle$ 
be two fragments that share the same ``authoritative'' source $u$, and a triple pattern $tp$.
If for all possible values of $\Gamma1$ and $\Gamma2$, 
always the triples in $\Gamma1$ that match $tp$ are also in $\Gamma2$, i.e.,
$\ldbrack tp \rdbrack_{\Gamma1}$ $\subseteq$ $\ldbrack tp \rdbrack_{\Gamma2}$.
Then, regarding $tp$, $f1$ is contained in $f2$. And we use the notation f1 $\sqsubseteq$ f2.
\label{def:containmentii}
\end{definition}
Triples of fragment \textit{f9} replicated at consumer \textit{C5} in Figure~\ref{fig:replicas} are contained in
fragment \textit{f1} at \textit{C1} and \textit{C3} (\textit{f9} $\sqsubseteq$ \textit{f1}) because  \textit{f1} and \textit{f9} share 
the same ``authoritative'' source, and all the triples
that match predicate \textit{p1} and object \textit{c1} always matches  predicate
\textit{p1} in \textit{f1}. Using fragment containment, contacting \textit{C1} or \textit{C3} is enough to answer \textit{Q1}.

\newsavebox{\querythree}
\begin{lrbox}{\querythree}
\begin{lstlisting}[basicstyle=\scriptsize\sffamily,language=sparql,numbers=none,frame=none,label=lst:queryc,breaklines=true,showstringspaces=false]
SELECT DISTINCT *
WHERE {
{?x1p1?x2.?x2p4?x3}
UNION
{?x1p2?x2.?x2p5?x3}
UNION
{?x1p3?x2.?x2p6?x3}
}
\end{lstlisting}
\end{lrbox}

\paragraph{Source Selection Problem (SSP)}
Given a set of SPARQL endpoints \textit{E}, a set of public endpoints \textit{P}, \textit{P} $\subseteq$ \textit{E},  the set of fragments contained in each endpoint as a
function \textit{frags : Endpoint $\rightarrow$ set of Fragment}, a
containment relation among endpoints (given by
Definition~\ref{def:containmenti}) for \textit{f $\in$ frags(e$_i$)
  $\land$ f $\in$ frags(e$_j$)}, \textit{e$_i$ $\subseteq_{f}$ e$_j$},
a containment relation among fragment selectors (given by
Definition~\ref{def:containmentii}) $f_l \sqsubseteq f_k$, and a SPARQL
query \textit{Q}. Find a map \textit{D}, such that for each triple
pattern \textit{tp} in \textit{Q},  \textit{D(tp) $\subseteq$ E} and:
\begin{inparaenum}[\itshape 1\upshape)]
  \item For each endpoint \textit{e} that may contribute with relevant data to answer query \textit{Q}, 
        \textit{e}  is included in \textit{D}, or \textit{D} includes another endpoint that contributes with at least 
        the same relevant data as \textit{e}.
  \item \textit{D(tp)} contains as few public endpoints as possible.
  \item \textit{size(D(tp))} is minimized for all triple pattern \textit{tp} in \textit{Q}.
  \item The number of different endpoints used within each basic graph pattern is minimized.  
\end{inparaenum}

\begin{figure}[t]
  \centering
\subfloat[][Query Q3]{\label{fig:ssp-q}
  \usebox{\querythree}
}
\subfloat[][RF for \textit{Q3}]{
\label{fig:ssp-rf}%
\begin{scriptsize}
\begin{tabular}{|c|c|c|}
 \hline
TP & RF & Endpoints \\ \hline
?x1 p1 ?x2 & f1 & C1, C3, P1\\
           & f9 & C5, P1 \\ \hline
?x2 p4 ?x3 & f4 & C2, C3, P1 \\ \hline
?x1 p2 ?x2 & f2 & C1, C4, P1\\ \hline
?x2 p5 ?x3 & f5 & C2, C4, P2\\ \hline
?x1 p3 ?x2 & f3 & C1, C5, P2\\ \hline
?x2 p6 ?x3 & f6 & C2, C5, P2\\ \hline
\end{tabular}
\end{scriptsize}
}
\subfloat[][Conditions 1-3]{%
\label{fig:ssp-d1d2}
\begin{scriptsize}
\begin{tabular}{|c|c|c|}
 \hline
tp & D$_1$(tp) & D$_2$(tp)\\ \hline
?x1 p1 ?x2 & \{ C1 \} & \{ C3 \}\\ \hline
?x2 p4 ?x3 & \{ C2 \} & \{ C3 \}\\ \hline
?x1 p2 ?x2 & \{ C1 \} & \{ C4 \}\\ \hline
?x2 p5 ?x3 & \{ C2 \} & \{ C4 \}\\ \hline
?x1 p3 ?x2 & \{ C1 \} & \{ C5 \}\\ \hline
?x2 p6 ?x3 & \{ C2 \} & \{ C5 \}\\  \hline
\end{tabular}
\end{scriptsize}
}
\subfloat[][Conditions 1-4]{\label{fig:ssp-d2}%
\begin{scriptsize}
\begin{tabular}{|c|c|c|}  \hline
BGP & tp          & D$_2$(tp) \\  \hline
BGP1 & ?x1 p1 ?x2 & \{ C3 \} \\
     & ?x2 p4 ?x3 & \{ C3 \} \\ \hline
BGP2 & ?x1 p2 ?x2 & \{ C4 \} \\
     & ?x2 p5 ?x3 & \{ C4 \} \\\hline
BGP3 & ?x1 p3 ?x2 & \{ C5 \} \\
     & ?x2 p6 ?x3 & \{ C5 \} \\
 \hline
\end{tabular}
\end{scriptsize}
}
\caption[]{SSP solutions for query Q3: \subref{fig:ssp-q}: Query Q3,
  \subref{fig:ssp-rf}:Relevant fragments for Q3,
  \subref{fig:ssp-d1d2}: Maps D$_1$ and D$_2$ that satisfy Source Selection Problem
(SSP) conditions 1-3., \subref{fig:ssp-d2}: Map D$_2$ that satisfies Source Selection Problem (SSP) conditions 1-4}
\label{fig:ssp}
\vspace{-0.5cm}
\end{figure}

Condition 1 states that the selected sources will produce an answer
as complete as possible given the set of fragments accessible through
the endpoints E, but answer may be incomplete if some fragments definitions are 
missing.
Condition 2 ensures that public endpoints
availability problem will be avoided whenever is possible. Condition 3
establishes that the number of selected sources is reduced. Condition 4
aims to reduce the size of intermediate results, and to delegate the join
execution to endpoints whenever is possible.
Even if the public endpoint can provide all the fragments, the 
use of several consumer endpoints is preferable.
To illustrate these four conditions, consider query \textit{Q3} 
in Figure~\ref{fig:ssp-q} and fragments and endpoints of
Figure~\ref{fig:replicas}. Table~\ref{fig:ssp-rf} shows the
relevant fragments for \textit{Q3} triple patterns, and the endpoints
that provide these fragments.
For example, for the triple pattern \textit{?x1 p1 ?x2}, there are two
relevant fragments $f1$ and $f9$. As previously discussed  using
endpoint containments, contacting \textit{C1} or \textit{C3} is enough to answer this triple pattern
without contacting the public endpoint.  The maps D$_1$ and
D$_2$ in Figure~\ref{fig:ssp-d1d2} satisfy the SSP conditions 1-3:
all  relevant fragments have been included directly or through containment relation,
 the number of selected endpoints per triple pattern has been minimized, and no public
endpoints has been included in the map. However, only the map D$_2$
satisfies condition 4, as the number of different endpoints
selected per basic graph pattern has been also minimized (see
Figure~\ref{fig:ssp-d2}). Then, joins are delegated to the selected 
endpoints, and the size of intermediate results is reduced. \emph{Current state-of-the-art~\cite{DBLP:conf/semweb/SaleemNPDH13} is triple pattern wise and does not
guarantee condition 4}.

\paragraph{Source Selection Algorithm}

Algorithm~\ref{sourceselectionalgorithm} sketches the \fedra source selection algorithm.
First, the algorithm pre-selects for each triple pattern in \textit{Q} the sources that can be used to
evaluate it (lines 2-29).  All the endpoints $e$ and their exposed fragments $f$ are considered (lines 5-27).
In line 6, the function {\it canAnswer()} is used to determine if endpoint $e$ can provide
triples from fragment $f$ that matches triple pattern $tp$. An initial check based on the selector of $f$ and $tp$ is
done, and when it is satisfied, a dynamic check using an ASK query is done to ensure that $f$ is relevant
for triple pattern $tp$ (as in Definition~\ref{def:fragmentrelevance}). An ASK query is used to avoid 
considering fragments that are not relevant for the triple pattern, in the case the triple pattern 
has constants where the fragment definition has variables.
Query $Q3$ relevant endpoints and fragments
are given  in Table~\ref{fig:ssp-rf}.


 \begin{algorithm} 
\scriptsize
 \caption{Source Selection algorithm}
 \label{sourceselectionalgorithm}
 \begin{multicols}{2}
\begin{algorithmic}[1]
\Require Q: SPARQL Query; E: set of Endpoints; frags : Endpoint $\rightarrow$ set of Fragment;
 P : set of Endpoint; $\subseteq_{f}$ : Endpoint $\times$ Endpoint; 
$\sqsubseteq$ : BGP $\times$ BGP 
\Ensure D: map from Triple Pattern to set of Endpoints.
\Function{sourceSelection}{Q,E,frags,P,$\subseteq_{f}$,$\subseteq$} 
\State triplePatterns $\leftarrow$ get triple patterns in Q
\For  {\textbf{each} tp $\in$ triplePatterns} 
 \State fragments $\leftarrow$ $\emptyset$
 \For {\textbf{each} e $\in$ E $\land$ f $\in$ frags(e)} 
   \If {canAnswer(e, f, tp)}
    \State include $\leftarrow$ true
    \For {\textbf{each} fs $\in$ fragments} 
     \State (f',e') $\leftarrow$ take one element of fs
     \If {subFrag(f,e,f',e',$\subseteq_{f}$,$\sqsubseteq$) $\land$ 
          subFrag(f',e',f,e,$\subseteq_{f}$,$\sqsubseteq$)}
     \State fs.add((f,e))
     \State include $\leftarrow$ false
     \Else \If {subFrag(f',e',f,e,$\subseteq_{f}$,$\sqsubseteq$)}
            \State fragments.remove(fs)
           \Else \If {subFrag(f,e,f',e',$\subseteq_{f}$,$\sqsubseteq$)}
                  \State include $\leftarrow$ false
                 \EndIf
           \EndIf
     \EndIf
    \EndFor
    \If {include}
     \State fragments.add(\{(f,e)\})
    \EndIf
   \EndIf
 \EndFor
 \State G(tp) $\leftarrow$ getEndpoints(fragments)
\EndFor \textbf{ each}
\State basicGP $\leftarrow$ get basic graph patterns in Q
\For {\textbf{each} bgp $\in$ basicGP}
 \State (S, C) $\leftarrow$ minimal set covering instance using bgp$\lhd$G
 \State C' $\leftarrow$ minimalSetCovering(S, C)
 \For  {\textbf{each} tp $\in$ bgp} 
  \State G(tp) $\leftarrow$ filter G(tp) according to C'
 \EndFor \textbf{ each}
\EndFor \textbf{ each}
\For  {\textbf{each} tp $\in$ domain(G)} 
 \State D(tp) $\leftarrow$ for each set in G(tp) include one element
\EndFor \textbf{ each}
\State \Return D
\EndFunction
\end{algorithmic}
\end{multicols}
\vspace{-0.25cm}
\end{algorithm}

The function {\it subFrag} determines if the data provided by one fragment is also provided by
another fragment. This function has as arguments the fragments and endpoints that provide them, and also the 
containment relationships. For each relevant fragment $f$, it determines if the considered fragment $f$ 
in endpoint $e$ provides the same data as already found fragments or if it
provides at least the same data as already found fragments or if it provides at most the same data as already
found fragments (lines 8-22). Function {\it subFrag} tests if there is a
containment in both senses or only in one of them.
Accordingly, the fragment is grouped with the fragments that provide the same data 
(lines 10-12), some of the already found fragments are not anymore of interest (line 14-15), 
 or it is chosen not to include the fragment (lines 17-18).
Between the fragments $f1$ and $f9$ selected as relevant fragments for the first triple pattern of $Q3$, there is a containment
$f9$ $\sqsubseteq$ $f1$, and in consequence, only the fragment $f1$ needs to be selected. 
Moreover, $f1$ is provided by endpoints
$C1$, $C3$, and $P1$, and as they provide the same data, each of them can be selected alone to provide
data for this triple pattern, i.e., they offer alternative sources for the same fragment.
Table~\ref{tab:fragmentConstruction} (left) shows the changes in $fragments$ for the first triple pattern
of $Q3$ .
Contrarily, between the fragments $f7$ and $f8$ selected as relevant for the second triple pattern of $Q2$,
there is no containment relation, and in consequence, both fragments are selected.
Table~\ref{tab:fragmentConstruction} (right) shows the changes in $fragments$ for the second triple pattern of $Q2$.

\begin{table}[t]
\caption{$fragments$ changes due to execution of lines 5-27 of 
Algorithm~\ref{sourceselectionalgorithm} for the first triple pattern of \textit{Q3} (left) and the 
second triple pattern of \textit{Q2} (right)} 
\label{tab:fragmentConstruction}
\begin{minipage}{0.4\linewidth}
\begin{table}[H]
\begin{center}
{\scriptsize \label{tab:fragmentConstructionA}
\begin{tabular}{|c|c|}
 \hline
line & fragments \\
 \hline
4    & \{ \} \\
\hline
24   & \{ \{ (f1, C1) \} \} \\
\hline
11  & \{ \{ (f1, C1), (f1, C3) \} \} \\
\hline
11  & \{ \{ (f1, C1), (f1, C3), (f1, P1) \} \} \\
\hline
\end{tabular}
}
\end{center}
\end{table}
\end{minipage}\hspace*{0.25cm}
\begin{minipage}{0.55\linewidth}
\begin{table}[H]
\begin{center}
{\scriptsize \label{tab:fragmentConstructionB}
\begin{tabular}{|c|c|}
 \hline
line & fragments \\
 \hline
4    & \{ \} \\
\hline
24   & \{ \{ (f7, C3) \} \} \\
\hline
24  & \{ \{ (f7, C3) \}, \{ (f8, C4) \} \} \\
\hline
11  & \{ \{ (f7, C3), (f7, P2) \}, \{ (f8, C4) \} \} \\
11  & \{ \{ (f7, C3), (f7, P2) \}, \{ (f8, C4), (f8, P2) \} \} \\
\hline
\end{tabular}
}
\end{center}
\end{table}
\end{minipage} 
\vspace{-0.85cm}
\end{table}

When the fragment provides non-redundant data, it is included in the selected fragments (lines 23-24). 
In the previous examples, it happens once for the  first triple pattern of \textit{Q3}
and twice for the second triple pattern of \textit{Q2}.
Function \textit{getEndpoints} in line 28 takes the endpoints of \textit{fragments}, and when 
for one subset \textit{fs} there are endpoints in \textit{E - P}, then all endpoints in \textit{P} are removed.
In the previous examples, the public endpoints are removed and 
the value of \textit{G(?x1 p1 ?x2)} is \textit{\{\{C1,C3\}\}}, and the value of \textit{G(?x1 p7 ?x3)} 
is \textit{\{\{ C3 \},\{C4\}\}}.
At the end of the first for loop (lines 3-29), a set whose elements are a set of 
endpoints  and fragments that can be used to evaluate the triple pattern
 is produced for each triple pattern in the query.
All the endpoints in the same set offer the same data for that fragment, and 
during execution only one of them needs to be contacted. 
And different elements of this resulting set correspond to different fragments that should be considered 
in order to obtain an answer as complete
as possible, modulo the considered fragments. 
For queries \textit{Q2} and \textit{Q3},  \textit{C1} and  \textit{C3} provide the same data for the first triple pattern of  \textit{Q3}, 
and \textit{C3}, \textit{C4} provide different data for the second  triple pattern of \textit{Q2} .

Next, for each basic graph pattern, a general selection takes place, considering the pre-selected sources for 
each triple pattern in the basic graph pattern. This last part can be reduced to the well-known
set covering problem, and an existing heuristic like the one given 
in~\cite{DBLP:books/daglib/0017733} may be used to perform the procedure 
indicated in line 33. To use the set covering heuristic, an instance of the set covering problem is produced
in line 32 using $bgp \lhd G$~\footnote{$bgp \lhd G$ represents function domain restriction, \emph{i.e.}, it takes the elements of
map \textit{G} that relates elements in \textit{bgp} with some set of set of endpoints}.
\begin{figure}[t]
 \centering
\subfloat[S instance]
{\label{fig:instance}
\begin{tikzpicture}[scale=0.7]
 \node at (0, 5) {\scriptsize tp1:};
 \node at (0, 4) {\scriptsize tp2:};
 \node at (1.25, 6) {\scriptsize Triple Pattern (tp)}; 
 \node at (1.5, 5) {\scriptsize ?x1 p4 ?x2};
 \node at (1.5, 4) {\scriptsize ?x1 p7 ?x3};
 \node at (4, 6) {\scriptsize G(tp)}; 
 \node at (4, 5) {\scriptsize \{\{C2,C3\}\}}; 
 \node at (4, 4) {\scriptsize \{\{C3\},\{C4\}\}}; 
 \node at (6.25, 6) {\scriptsize S}; 
 \node at (6, 5) {\scriptsize \{ s$_{1,1}$,}; 
 \node at (6.4, 4) {\scriptsize    s$_{2,1}$, s$_{2,2}$ \}};
 \draw [thick, <-] (6.15,5.3) -- (6.15,5.5) -- (4, 5.5) -- (4, 5.3); 
 \draw [thick, <-] (5.8,4.2) -- (5.8,4.5) -- (3.6, 4.5) -- (3.6, 4.2); 
 \draw [thick, <-] (6.6,3.7) -- (6.6,3.4) -- (4.4, 3.4) -- (4.4, 3.7);   
\end{tikzpicture}
}\hspace*{1cm}
\subfloat[C instance]
{\label{fig:collection}
\begin{tikzpicture}[scale=0.7]
 \node at (0, 0) {\scriptsize \{\{s$_{1,1}$\},\{s$_{1,1}$,s$_{2,1}$\},\{s$_{2,2}$\}\} };
 \node at (-1.65, 1) {\scriptsize C2};
 \node at (0, 1) {\scriptsize C3}; 
 \node at (1.75, 1) {\scriptsize C4};
 \draw [thick, ->] (-1.65,0.75) -- (-1.65, 0.5); 
 \draw [thick, ->] (0,0.75) -- (0, 0.5); 
 \draw [thick, ->] (1.75,0.75) -- (1.75, 0.5);   
\end{tikzpicture}
}
 \caption{Set covering instances of S and C for the query $Q2$ and federation given in Figure~\ref{fig:replicas}. 
 For each element in G(tp), one element is included in set S. For each endpoint  in G, one set is included in collection C and its elements
 are the elements of S related to the endpoint}
 \vspace{-0.5cm}
\end{figure}
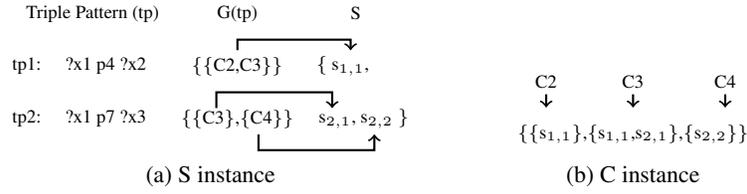
Figure~\ref{fig:instance} shows \textit{G} values obtained after lines 3-29 loop has ended
for query \textit{Q2}, and S instance for the set covering problem. For each set in \textit{G(tp)}, 
one element is included in \textit{S}, e.g., for set \textit{\{C2,C3\}}, the element \textit{s$_{1,1}$} 
is included in \textit{S}.
We have used subscripts \textit{i,j} to denote that the element comes from the triple pattern \textit{i}, 
and it is the \textit{j}-th element coming from this triple pattern.
The collection \textit{C} is composed of one set for each endpoint that is present in \textit{G}, 
and its elements are the elements of \textit{S} related to each endpoint. 
Figure~\ref{fig:collection} shows the instance of \textit{C} for this example.
The collection \textit{C'} obtained in line 33 is \textit{\{C3,C4\}}, as the union of \textit{C3} 
and  \textit{C4} is equals to \textit{S}, and there is no smaller subset of \textit{C} that can achieve this.
The instruction  in line 35 removes from each  \textit{G(tp)} the elements that do not belong to  \textit{C'}.
In the example,  \textit{C2} is removed from  \textit{G(?x1 p4 ?x2)}.
A last step may be performed to choose among endpoints that provide the same fragment and ensure 
a better query processing by existing federated query engines (lines 38-40).
Nevertheless, these alternative sources could be used to speed up query processing, 
e.g., by getting a part of the answer from each endpoint.

\section{Experiments}
\label{sec:experiments}

We conducted many experimentations in different setups to demonstrate
the impact of \fedra on existing approaches, complete results are
available at the \fedra web
site\footnote{\url{https://sites.google.com/site/fedrasourceselection}}. The
performance of the FedX and the ANAPSID query engines 
 is our baseline. 
 General
results are the comparison of performance of FedX alone, FedX+DAW,
FedX+\fedra, same thing for ANAPSID. Compared to FedX and ANAPSID, 
\fedra should reduce selected sources significantly and speed up
queries execution time. Compared to DAW, we expect \fedra to
achieve same source reduction but without pre-computed MIPS index, and
generate less intermediate results thanks to endpoints reduction 
and to finding join opportunities.

\newcommand{\datasetstab}{\scriptsize
\begin{tabular}{|c|c|c|c|}
 \hline
Dataset & Version date & \# DT & \# P \\
\hline
Diseasome               & 19/10/2012 & 72,445 & 19\\
Semantic Web Dog Food   & 08/11/2012 & 198,797 & 147\\
DBpedia Geo-coordinates & 06/2012    & 1,900,004 & 4\\
LinkedMDB               & 18/05/2010 & 3,579,610 & 148\\
\hline
\end{tabular}
}

\newcommand{\queriesnumtab}{\scriptsize
\begin{tabular}{|c|c|c|c|c|c|c|}
 \hline
Dataset                 & ST & 2P & 3P & 4P & 2S & 3S\\
\hline
Diseasome               & 5 & 4 & 5 & 2 & 5 & 5 \\
Semantic Web Dog Food   & 5 & 7 & 7 & 4 & 5 & 5 \\
DBpedia Geo-coordinates & 5 & 0 & 0 & 0 & 5 & 5 \\
LinkedMDB               & 5 & 0 & 0 & 0 & 0 & 0 \\
\hline
\end{tabular}
}

\newcommand{\querieschartab}{\scriptsize
\begin{tabular}{|c|c|c|c|c|c|c|c|}
 \hline
Dataset                 & J & OP & U & F(R) & L & OB\\
\hline
Diseasome               & 19 & 2 & 1 & 11(7) & 4 & 9\\
Semantic Web Dog Food   & 24 & 5 & 2 & 8(7) & 4 & 9\\
DBpedia Geo-coordinates & 10 & 0 & 0 & 6(1) & 6 & 6\\
LinkedMDB               & 0 & 0 & 0 & 1(1) & 1 & 2\\
\hline
\end{tabular}
}

\newcommand{\federations}{\scriptsize
\begin{tabular}{|c|c|c|c|}
 \hline
Dataset                 & \# FD & \# E1F & \# E2F\\
\hline
Diseasome               & 16 & 16 & 120 \\
Semantic Web Dog Food   & 40 & 40 & 780 \\
DBpedia Geo-coordinates & 4 & 4 & 6 \\
LinkedMDB               & 4 & 4 & 6 \\
\hline
\end{tabular}
}

\newcommand{\precomputation}{\scriptsize
\begin{tabular}{|c|c|c|}
 \hline
Federation              & DAW Index & \fedra Containment\\
\hline
Diseasome               & 199.69 & 14,605\\
Semantic Web Dog Food   & 536.21 & 114,340\\
DBpedia Geo-coordinates & 8,778.03 & 28\\
LinkedMDB               & 539.1 & 14\\
\hline
\end{tabular}
}

\begin{table}[t]
\centering
\subfloat[Datasets]{\label{tab:datasets}\datasetstab}
\subfloat[Queries sizes and number]{\label{tab:queriesnum}\queriesnumtab}\\
\subfloat[Queries with operators, modifiers]{\label{tab:querieschar}\querieschartab}
\subfloat[Federations]{\label{tab:federations}\federations}
\caption{Datasets, queries and federations characteristics. For the datasets: the version, number of different triples (\# DT), 
and predicates (\# P). For the queries: the number of queries with 1 triple pattern (ST), 
2, 3 or 4 triple patterns in star shape (S) or path shape (P). Also the number of queries with: joins (J),
optionals (OP), unions (U), filter (F), regex expressions (R), limit (L), and order by (OB). 
For the federations: the number of fragments definitions (FD), 
endpoints exposing one and two fragments (E1F, E2F)
}
\vspace{-0.75cm}
\end{table}

\noindent {\bf Datasets, Queries and Federations Benchmark:} we used 
Diseasome, Semantic Web Dog Food, LinkedMDB, and DBpedia
geo-coordinates datasets, Table~\ref{tab:datasets} shows characteristics of the
evaluated datasets.  We studied the datasets and queries used
in~\cite{DBLP:conf/semweb/SaleemNPDH13}\footnote{They are available
  at~\url{https://sites.google.com/site/dawfederation}.}. However, we
modified the queries to include the DISTINCT modifier in all the queries.
Additionally, the ORDER BY clause was included in the queries with the LIMIT clause, in order to make them
susceptible to a reduction in the set of selected sources without
changing the query answer, and to ensure a semantically unambiguous query
answer.  Tables~\ref{tab:queriesnum} and \ref{tab:querieschar}
present queries characteristics.  As federations used
in~\cite{DBLP:conf/semweb/SaleemNPDH13} do not take into account
fragments, they were not reprised. A federation was set up for each
dataset; each federation is composed of the triple pattern fragments
that are relevant for the studied queries.  The federation endpoints
 offer one or two fragments; endpoints with two fragments
offer opportunities to execute joins for the
engine.  Table~\ref{tab:federations} shows the federation
characteristics. 
In average, DAW indexes were computed in 2,513 secs, and \fedra containments in 32 secs.

Notice that as \fedra containments depends only on fragment descriptions, 
their updates are less frequent than DAW indexes.
%
Virtuoso 6.1.7\footnote{\url{http://virtuoso.openlinksw.com/}, November 2013.} endpoints were used.
A Virtuoso server was set up to provide all the endpoints as virtual endpoints. It was
configured with timeouts of 600 secs. and 100,000 tuples. 
In order to measure the size of intermediate results, proxies were used to access the endpoints.

\def\lstlistingname{Listing}

 \noindent
{\bf Implementations:} FedX 3.0\footnote{\url{http://www.fluidops.com/fedx/}, September 2014.}
 and ANAPSID\footnote{\url{https://github.com/anapsid/anapsid}, September 2014.} have been 
 modified to call \fedra and DAW~\cite{DBLP:conf/semweb/SaleemNPDH13} source selection
strategies during  query processing. Thus, each engine can use the selected
sources to perform its own optimization strategies.
Because FedX  is implemented in Java, while ANAPSID is implemented in Python,  
\fedra and DAW\footnote{We had to implement DAW as its code is not available.} were  
implemented in both Java 1.7 and Python 2.7.3.. Thus,  \fedra and DAW were integrated in FedX and ANAPSID, 
reducing the performance impact of including these new source selection strategies.
Proxies were implemented in Java 1.7. using the Apache HttpComponents Client library 
4.3.5.\footnote{\url{https://hc.apache.org/}, October 2014.}

\noindent
{\bf Evaluation Metrics:}
\begin{inparaenum}[\itshape i\upshape)]
 \item {\it Number of Selected Public Sources (NSPS):} is the sum of the number of 
 public sources that has been selected per triple pattern.
 \item {\it Number of Selected Sources (NSS):} is the sum of the number of 
 sources that has been selected per triple pattern.
 \item {\it Execution Time (ET):} is the elapsed time since the query is posed 
 by the user and the answers are completely produced. It is detailed in source selection 
 time (SST), and total execution time (TET). 
 Time is expressed in seconds (secs.). A timeout of 300 secs. has been enforced. 
 Time was measured using the bash time command.
 \item {\it Intermediate Results (IR):} is the number of tuples transferred from 
 all the endpoints to the query engine during a query evaluation. 
 \item {\it Recall (R):} is the proportion of results obtained by the underlying 
 engine, that are also obtained including proposed strategy.
\end{inparaenum}

\paragraph{Impact of the number of containments over FEDRA behavior}

 \begin{figure}[t]
 \centering
 {\includegraphics[width=1.05\textwidth,height=4.3cm]{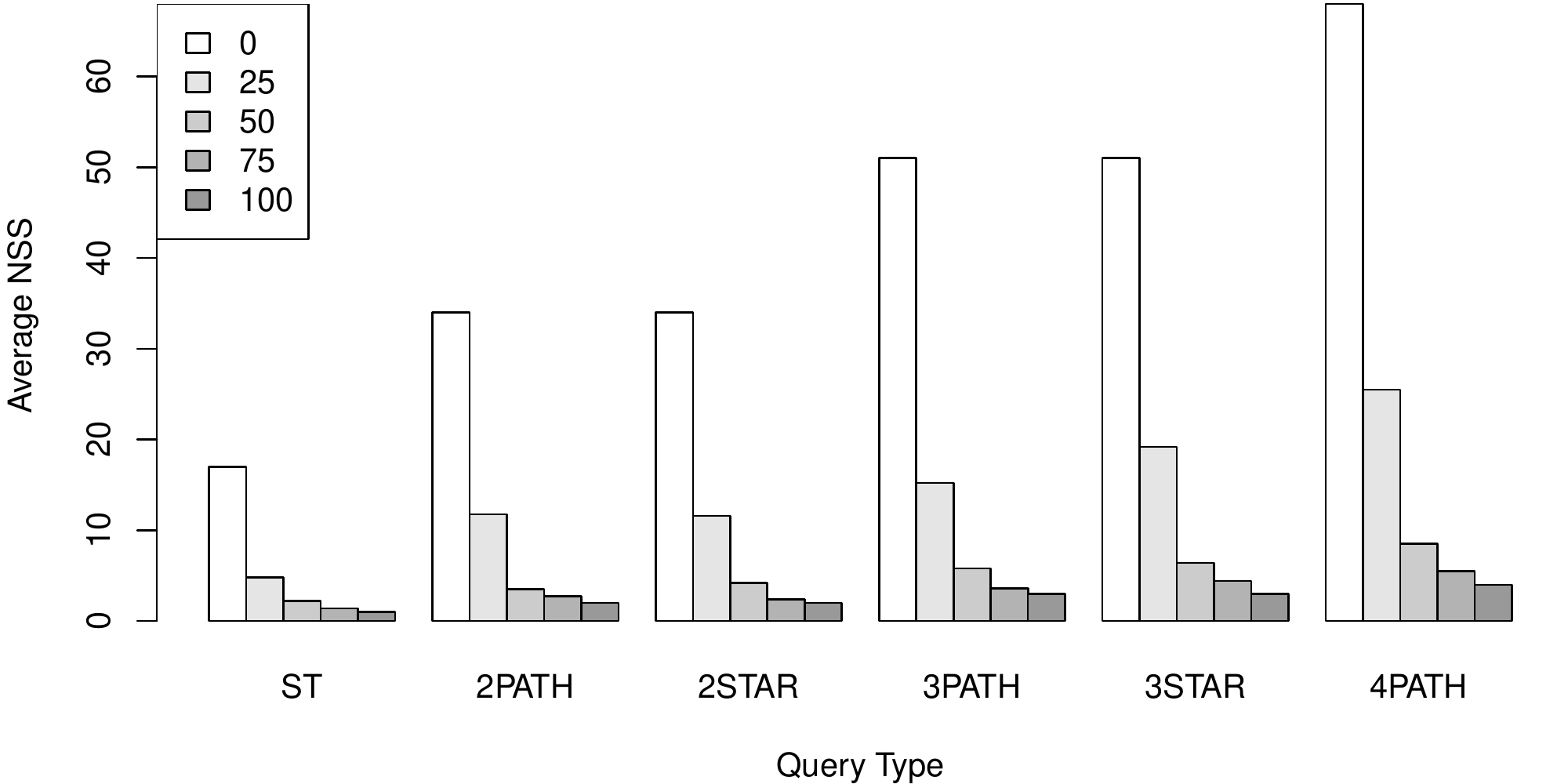}}\\
 \caption{FedX and \fedra Number of Selected Sources  (NSS) when the percentage of known containments is 0\%, 25\%, 50\%, 75\% and 100\%, for Diseasome Federation}
 \label{fig:numberSelectedSources}
 \vspace{-0.25cm}
 \end{figure}
 

We study the impact of the number of containments known during the source selection on 
the number of selected sources and intermediate results size.  
For each query, the set of known containments have been 
set to contain 0\%, 25\%, 50\%, 75\%, and 100\% of the containments concerning the relevant fragments,
and queries have been executed in these five configurations.
Results show that the number of selected public sources is equal to the number of triple patterns in 
the queries when no containment is known. However, as soon as some containments are known (25\%-100\%), this number is
reduced to zero. Also, the number of selected sources is considerably
reduced when \fedra source selection strategy is used instead of 
just using ANAPSID and FedX source selection strategies; see FedX results in Figure~\ref{fig:numberSelectedSources}. 
ANAPSID results exhibit a similar behavior.

\paragraph{Preservation of the Query Answer}
The goal of this experiment is to determine the impact of \fedra source selection strategy on query completeness. 
Queries were executed  using both the ANAPSID and the FedX query engines, and then, we executed the same engines enhanced with the
\fedra source selection strategy. Recall was computed and was 1.0  in the majority of the cases.
In few cases the recall was considerably reduced, but these cases correspond to queries with OPTIONAL operator using
the FedX query engine, and it was due to an implementation error for this operator.
\fedra only discards relevant sources when relevant fragments data are provided by another source that 
was already selected. Then, it does not reduce the recall of the answer.
Finally,  the query engine implementation limitations were also the causes of the reduction of recall.


\paragraph{Source Selection Time}

\begin{figure}[t]
 \centering
 \subfloat{\label{fig:sourceSelectionTimeOne}\includegraphics[width=0.5\textwidth,height=3.5cm]{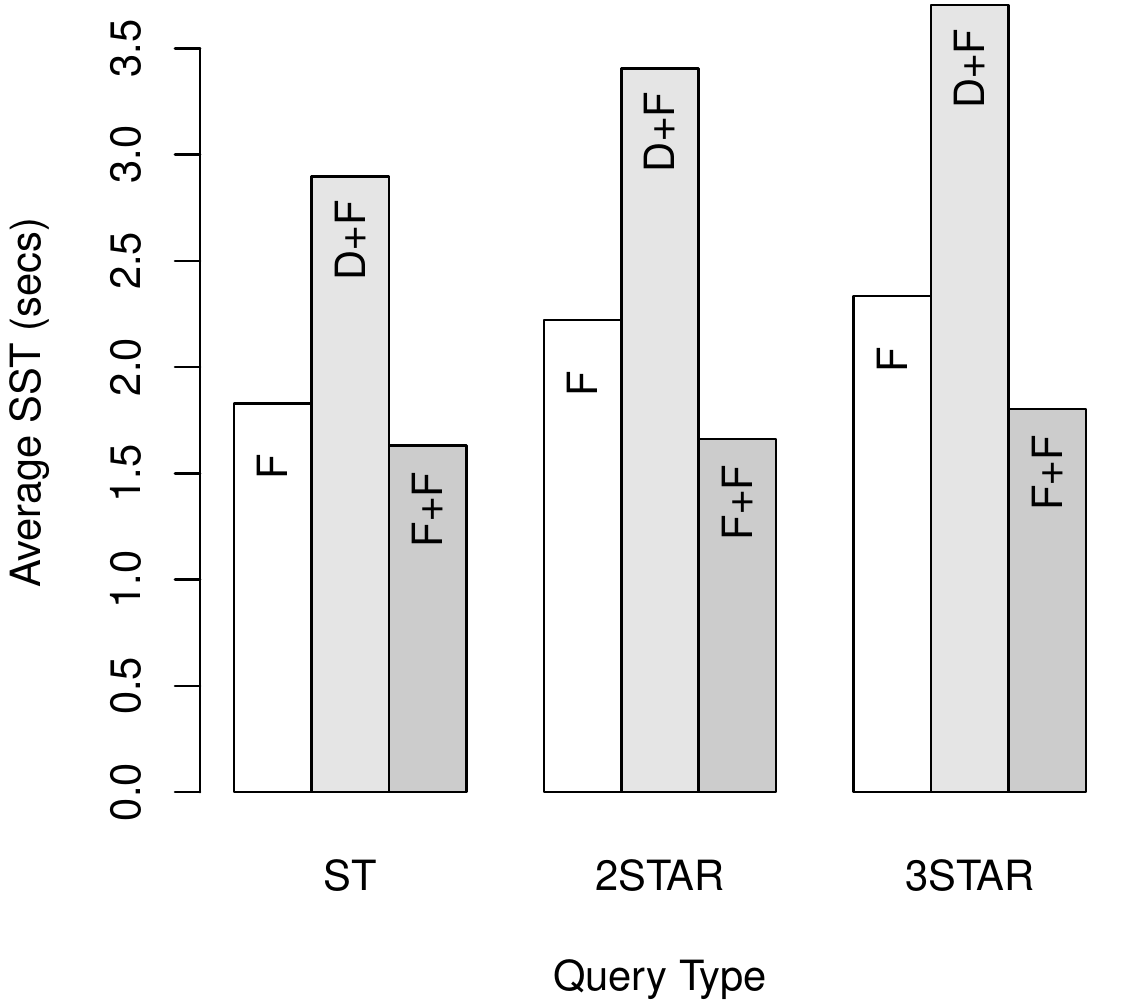}}
 \subfloat{\label{fig:sourceSelectionTimeTwo}\includegraphics[width=0.5\textwidth,height=3.5cm]{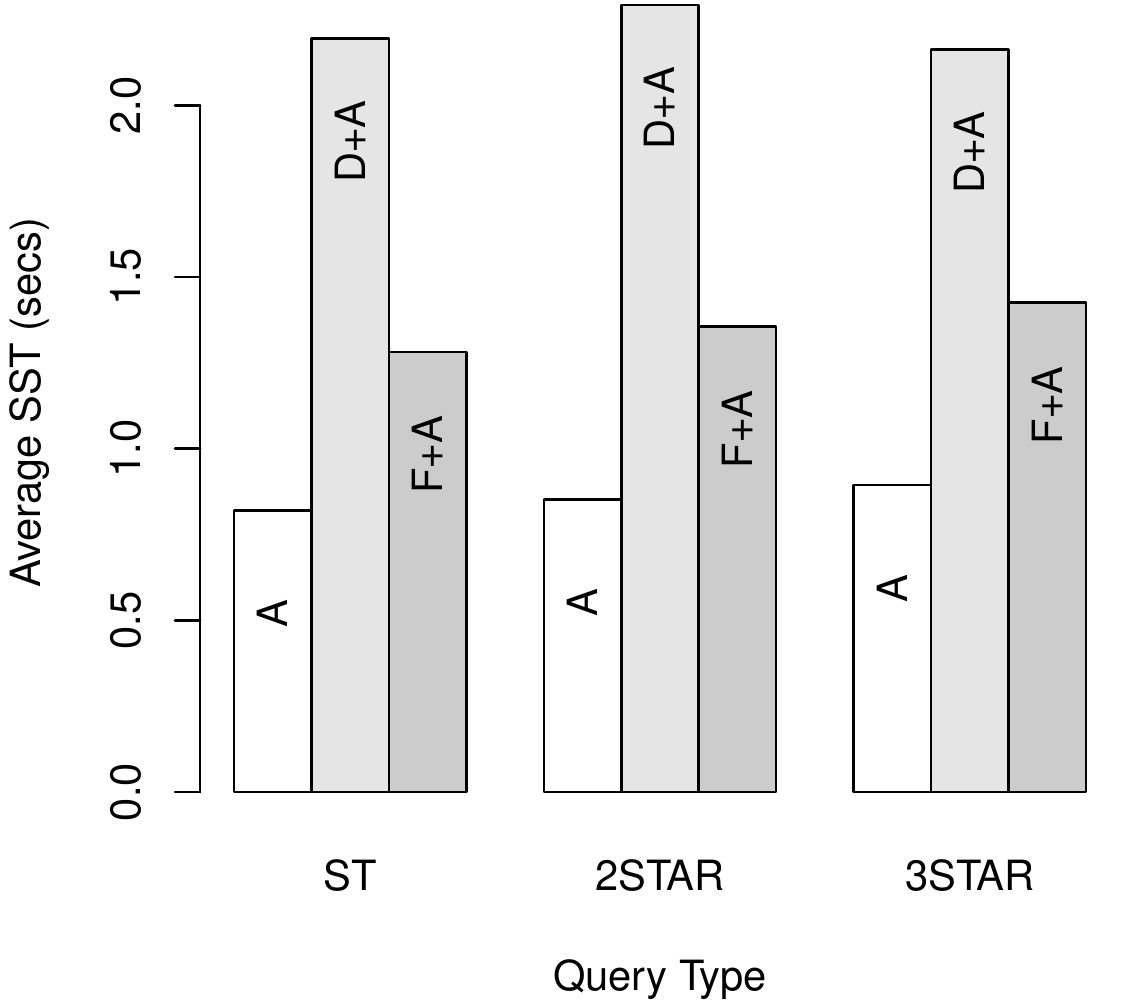}}\\
 \caption{Source selection time (SST) for Geo-coordinates federation. The FedX (F, left) and the ANAPSID (A, right) query
 engines are combined with \fedra (F+F and F+A), and DAW (D+F and D+A)}
 \vspace{-0.25cm}
 \end{figure} 

To measure the \fedra and DAW source selection cost, the source selection time using each engine with 
and without \fedra or DAW was measured. Results are diverse, for federations with a large number of endpoints
like the SWDF federation, the cost of performing the \fedra source selection is considerably inferior
to the cost of contacting all the endpoints using FedX, but similar to the cost of using the ANAPSID 
source selection strategy. On the other hand, the DAW cost is similar to the FedX cost, and it is considerably superior to the 
ANAPSID cost.
For federations with a small number of endpoints like Geo-coordinates (see Figures~\ref{fig:sourceSelectionTimeOne} 
and~\ref{fig:sourceSelectionTimeTwo}), the cost of performing source selection with \fedra is less expensive
than performing just FedX source selection, but more expensive than performing just ANAPSID source selection strategy. And the cost of performing source selection with DAW is more expensive than performing just FedX or
ANAPSID source selection strategy. 
ANAPSID source selection mostly relies on the endpoints descriptions, and avoids to contact endpoints in most
cases. On the other hand, FedX source selection strategy relies on endpoint contacts to determine
which endpoints can be used to obtain data for each triple pattern. \fedra source selection is somewhere
in the middle, it does contact all the endpoints that are considered relevant according to their descriptions to
confirm that they can provide relevant data, and has the added cost of using the containments
to reduce the number of selected sources. DAW source selection does not contact sources, but relies on no 
negligible cost of operating Min-Wise Independent Permutations (MIPs) in order to determine the overlapping sources.
 
\paragraph{Execution Time}

\begin{figure}[t]
 \centering
 \subfloat{\label{fig:numberSelectedSourcesTotal}\includegraphics[width=0.5\textwidth,height=3.5cm]{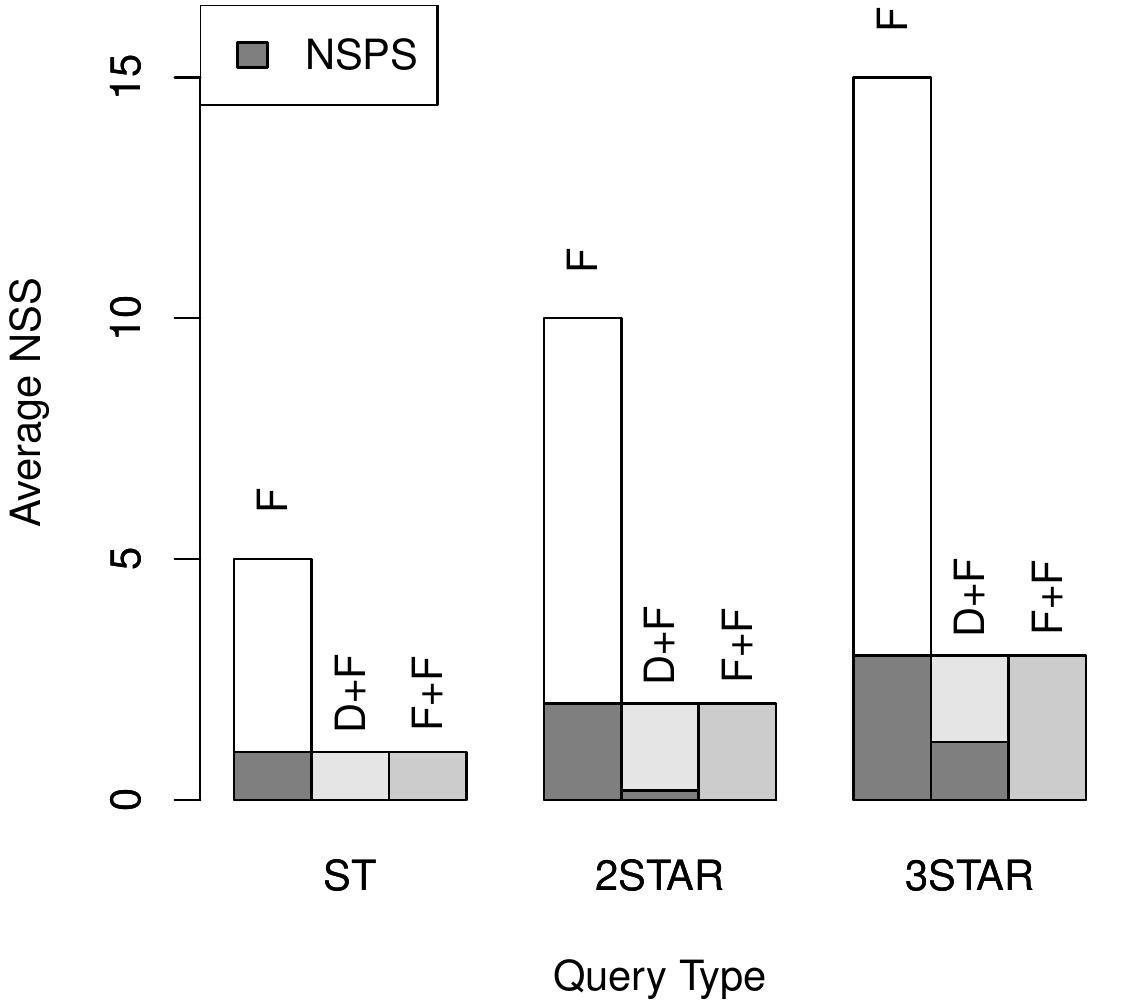}}
 \subfloat{\label{fig:executionTime}\includegraphics[width=0.5\textwidth,height=3.5cm]{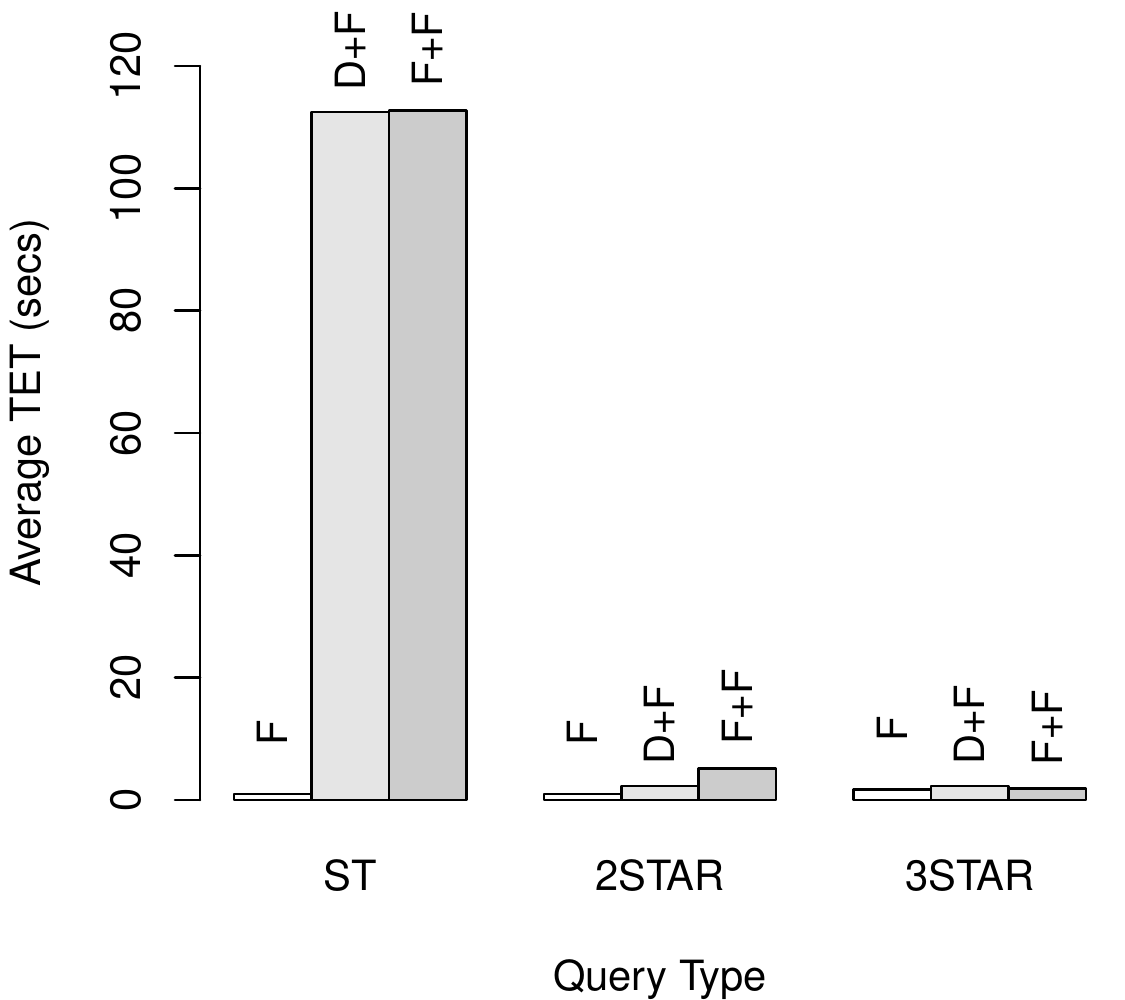}}
 \caption{Number of Selected Sources and Total Execution Time (TET) for Geo-coordinates federation and the FedX (F) query engine. 
 FedX is also combined with \fedra and DAW source selection strategies (F+F and D+F). For F, 4 out of 5 queries timed out for each query type.
 For F+F, 4 out of 5 queries timed out for 3STAR queries. For D+F, 4 out of 5 queries timed out for 2STAR and 3STAR queries}
 \label{fig:numberSelectedSourcesTwo}
 \vspace{-0.3cm}
 \end{figure}

To measure the \fedra and DAW execution time, queries were executed using each engine with 
and without \fedra or DAW. For small federations or queries with only one triple pattern, \fedra 
and DAW achieve a similar reduction in execution time. However, for larger federations and queries with more
triple patterns, \fedra reduction is larger than DAW. In all cases, the reduction is considerable when the combination of \fedra
and the query engine is compared to using the engine alone, e.g., Figure~\ref{fig:executionTime} shows the results 
for the Geo-coordinates federation and FedX. For star-shape queries, the use of \fedra source
selection strategy has made the difference between timing out or obtaining answers. For queries with two triple patterns, this difference is important,
 as \fedra enhances FedX to obtain answers in just few seconds.
The difference in execution time is a direct consequence of the selected sources reduction. Further,  
 executing the joins in the endpoints whenever it is possible, may reduce the size of
intermediate results and produce answers sooner.

\paragraph{Reduction of the Number of Selected Sources}

To measure the reduction of the number of selected sources, the source selection was performed
using ANAPSID and FedX with and without \fedra or DAW. For each query, the sum of the number of selected
sources per triple pattern was computed, for all the sources and just for the public sources. 
Figure~\ref{fig:numberSelectedSourcesTwo} shows the results for the
Geo-coordinates federation and FedX, similar results are observed for the other federations and for ANAPSID.
DAW source selection strategy exhibits the same reduction in the total number of selected sources.
Consequently, some of the selected public sources are pruned, but as it does not aim to reduce 
the public sources, it does not achieve a consistent reduction of them. On the other hand, \fedra 
has as input the public condition of sources, and as one of its goals is to select as few public sources as
possible, it is natural to observe such a reduction consistently for all the query types.
\fedra source selection strategy identifies the relevant fragments and endpoints that provide 
the same data. Only one of them is actually selected, and in consequence, a huge reduction on the number 
of selected sources is achieved. Moreover, 
public endpoints are safely removed from the selected sources as their data can be retrieved from other sources.

\paragraph{Reduction of the Intermediate Results Size}

\begin{figure}[t]
 \centering
\subfloat[Federation with public endpoint]{\label{fig:intermediateResultsSizeA}
 \includegraphics[width=0.52\textwidth,height=4.2cm]{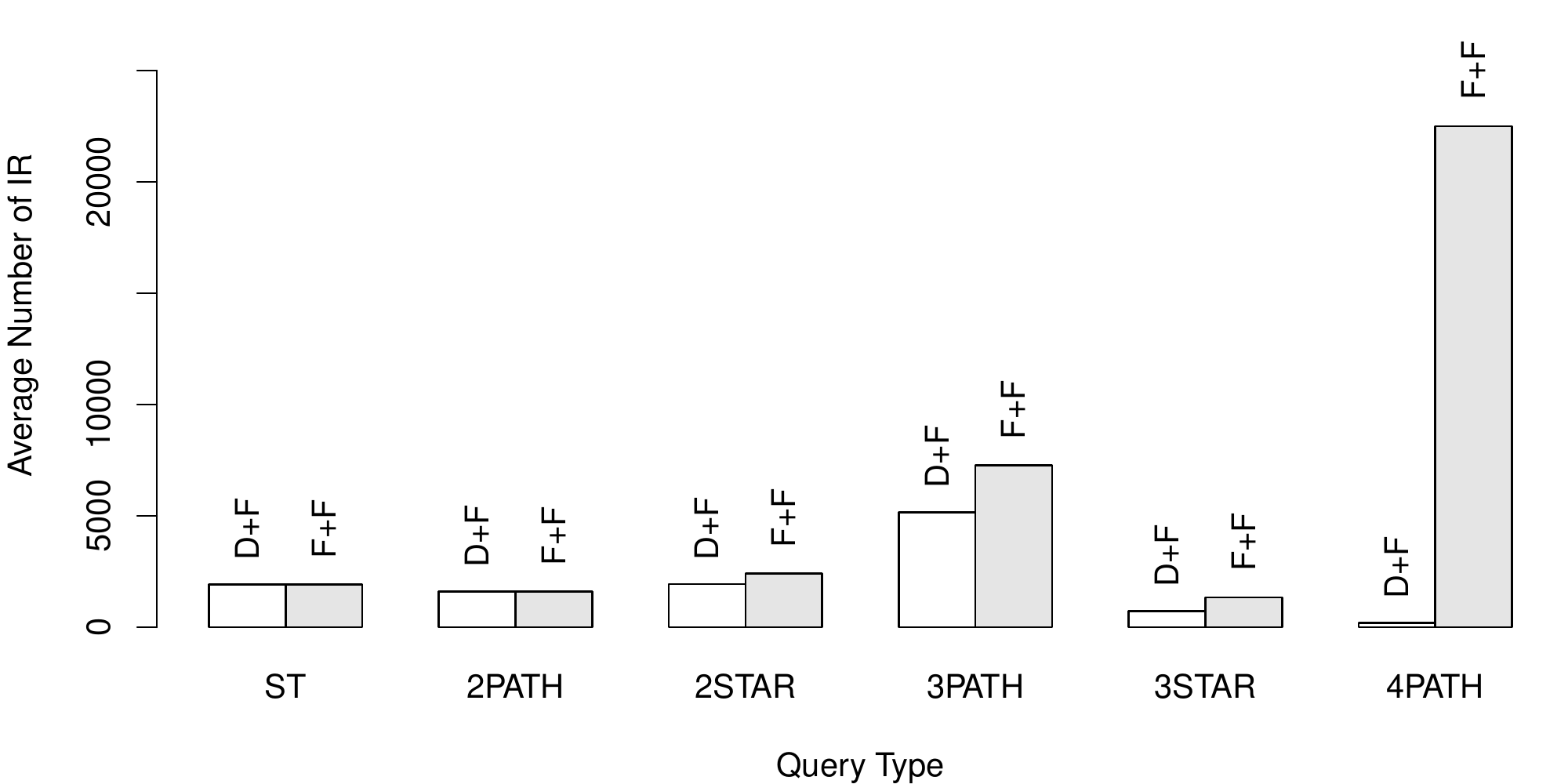}}
\subfloat[Federation without public endpoint]{\label{fig:intermediateResultsSizeB}
 \includegraphics[width=0.52\textwidth,height=4.2cm]{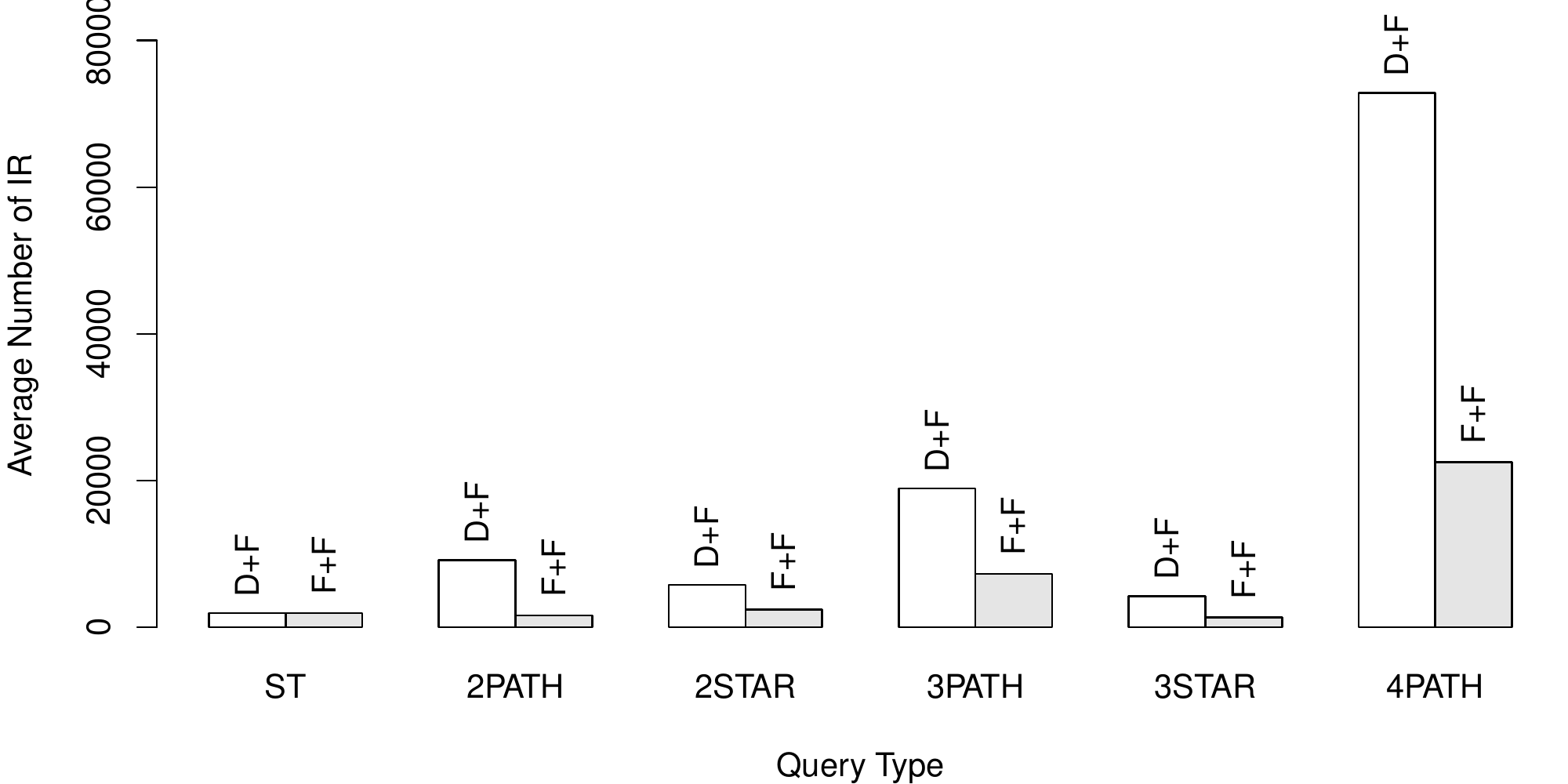}} 
 \caption{Intermediate results size for Diseasome federation and FedX (F). FedX is also combined with \fedra and DAW source selection strategies (F+F and D+F)}
 \label{fig:intermediateResultsSizeTwo}
 \vspace{-0.3cm}
 \end{figure}
 
To measure the intermediate results size reduction, queries were executed using proxies that measure the
number of transmitted tuples from endpoints to the engines.
Additionally, each query was executed against the federation with and without the public endpoint.
Figure~\ref{fig:intermediateResultsSizeTwo} shows the sizes of intermediate results for the Diseasome federation
and FedX combined with \fedra and DAW; similar results were obtained for the other federations and for ANAPSID. 
Figure~\ref{fig:intermediateResultsSizeA} shows that when the public endpoint is part of the federation, DAW
source selection strategy leads to executions with considerably less intermediate results. 

Figure~\ref{fig:intermediateResultsSizeB} shows that when the public 
endpoint is not part of the federation, \fedra source selection strategy  leads to executions with
considerably less intermediate results.
Since the \fedra source selection strategy finds opportunities to execute joins in the endpoints, 
and mostly, it leads to significant reduction in the intermediate results size. These results are 
consequence of SSP Condition 4, and cannot be systematically achieved by DAW as it is a triple 
wise based approach.
Nevertheless, as DAW source selection does not avoid  public endpoints,   it may select to execute all 
triple patterns in the public endpoint, and this comes with a huge reduction of the size of intermediate results.
Figure~\ref{fig:intermediateResultsSizeB} shows that when this ``public endpoint'' opportunity to execute all triple
patterns in one endpoint is removed, DAW source selection strategy does not consistently reduce the 
intermediate results size.


\section{Conclusions and Future Works}
\label{sec:conclusion}

Recent works on replicated fragments spread data and SPARQL processing
capabilities over data consumers. This opens new opportunities for
federated query processing by offering new tradeoffs between
availability and performance.
We presented \fedra, a source selection approach that takes advantage
of replicated fragment definitions to reduce the use of public
endpoints as they are mostly overloaded.  \fedra identifies and uses
opportunities to perform join in the sources relieving the query
engine of performing them, and reducing the size of intermediate
results.  
Experimental results demonstrate that the number of selected sources
remains low even when high number of endpoints and replicated fragments
are part of the federations. 
Results rely on containments induced by
replicated fragment definitions. Next, selecting the same sources for
Basic Graph Patterns triples is a strategy that allows
to reduce significantly the number of intermediate results.
Perspectives include dynamic discovery of endpoints providing
replicated fragments. This allows federated query engines to expand at
run-time declared federations with consumer endpoints of
interests. Such mechanism can improve both data availability and
performances of federated queries in Linked Data.

\bibliographystyle{abbrv}
\bibliography{references}

\begin{thebibliography}{10}

\bibitem{DBLP:conf/semweb/AcostaVLCR11}
M.~Acosta, M.-E. Vidal, T.~Lampo, J.~Castillo, and E.~Ruckhaus.
\newblock Anapsid: An adaptive query processing engine for sparql endpoints.
\newblock In {\em ISWC}, pages 18--34, 2011.

\bibitem{DBLP:conf/semweb/ArandaHUV13}
C.~B. Aranda, A.~Hogan, J.~Umbrich, and P.-Y. Vandenbussche.
\newblock Sparql web-querying infrastructure: Ready for action?
\newblock In {\em ISWC (2)}, pages 277--293, 2013.

\bibitem{DBLP:conf/semweb/BascaB10}
C.~Basca and A.~Bernstein.
\newblock Avalanche: Putting the spirit of the web back into semantic web
  querying.
\newblock In A.~Polleres and H.~Chen, editors, {\em ISWC Posters{\&}Demos},
  volume 658 of {\em CEUR Workshop Proceedings}. CEUR-WS.org, 2010.

\bibitem{DBLP:journals/ijswis/BizerHB09}
C.~Bizer, T.~Heath, and T.~Berners{-}Lee.
\newblock Linked data - the story so far.
\newblock {\em IJSWIS}, 5(3):1--22, 2009.

\bibitem{DBLP:journals/jcss/BroderCFM00}
A.~Z. Broder, M.~Charikar, A.~M. Frieze, and M.~Mitzenmacher.
\newblock Min-wise independent permutations.
\newblock {\em JCSS}, 60(3):630--659, 2000.

\bibitem{DBLP:books/daglib/0017733}
S.~Dasgupta, C.~H. Papadimitriou, and U.~V. Vazirani.
\newblock {\em Algorithms}.
\newblock McGraw-Hill, 2008.

\bibitem{DBLP:conf/semweb/GorlitzS11}
O.~G{\"o}rlitz and S.~Staab.
\newblock Splendid: Sparql endpoint federation exploiting void descriptions.
\newblock In O.~Hartig, A.~Harth, and J.~Sequeda, editors, {\em COLD}, 2011.

\bibitem{DBLP:conf/sigmod/HoseS12}
K.~Hose and R.~Schenkel.
\newblock Towards benefit-based {RDF} source selection for {SPARQL} queries.
\newblock In {\em SWIM}, page~2, 2012.

\bibitem{colgraph}
L.-D. Ibáñez, H.~Skaf-Molli, P.~Molli, and O.~Corby.
\newblock Col-graph: Towards writable and scalable linked open data.
\newblock In {\em ISWC}, 2014.

\bibitem{DBLP:journals/csur/Kossmann00}
D.~Kossmann.
\newblock The state of the art in distributed query processing.
\newblock {\em {ACM} Computer Survey}, 32(4):422--469, 2000.

\bibitem{Fedra}
G.~Montoya, H.~Skaf-Molli, P.~Molli, and M.-E. Vidal.
\newblock {Fedra: Query Processing for SPARQL Federations with Divergence}.
\newblock Technical report, Universit{\'e} de Nantes, May 2014.

\bibitem{ozsu2011principles}
M.~T. {\"O}zsu and P.~Valduriez.
\newblock {\em Principles of distributed database systems}.
\newblock Springer, 2011.

\bibitem{DBLP:journals/tods/PerezAG09}
J.~P{\'{e}}rez, M.~Arenas, and C.~Gutierrez.
\newblock Semantics and complexity of {SPARQL}.
\newblock {\em {ACM} TODS}, 34(3), 2009.

\bibitem{DBLP:conf/esws/QuilitzL08}
B.~Quilitz and U.~Leser.
\newblock Querying distributed {RDF} data sources with {SPARQL}.
\newblock In {\em ESWC}, pages 524--538, 2008.

\bibitem{DBLP:conf/semweb/SaleemNPDH13}
M.~Saleem, A.-C.~N. Ngomo, J.~X. Parreira, H.~F. Deus, and M.~Hauswirth.
\newblock Daw: Duplicate-aware federated query processing over the web of data.
\newblock In {\em ISWC}, pages 574--590, 2013.

\bibitem{DBLP:conf/esws/SaleemN14}
M.~Saleem and A.~N. Ngomo.
\newblock Hibiscus: Hypergraph-based source selection for {SPARQL} endpoint
  federation.
\newblock In {\em ESWC}, pages 176--191, 2014.

\bibitem{DBLP:conf/semweb/SchwarteHHSS11}
A.~Schwarte, P.~Haase, K.~Hose, R.~Schenkel, and M.~Schmidt.
\newblock Fedx: Optimization techniques for federated query processing on
  linked data.
\newblock In {\em ISWC}, pages 601--616, 2011.

\bibitem{DBLP:journals/www/UmbrichHKHP11}
J.~Umbrich, K.~Hose, M.~Karnstedt, A.~Harth, and A.~Polleres.
\newblock Comparing data summaries for processing live queries over linked
  data.
\newblock {\em WWW}, 14(5-6):495--544, 2011.

\bibitem{verborgh_iswc_2014}
R.~Verborgh, O.~Hartig, B.~De~Meester, G.~Haesendonck, L.~De~Vocht,
  M.~Vander~Sande, R.~Cyganiak, P.~Colpaert, E.~Mannens, and R.~Van~de Walle.
\newblock Querying datasets on the {Web} with high availability.
\newblock In {\em ISWC}, 2014.

\bibitem{verborgh_ldow_2014}
R.~Verborgh, M.~Vander~Sande, P.~Colpaert, S.~Coppens, E.~Mannens, and
  R.~Van~de Walle.
\newblock Web-scale querying through {Linked Data Fragments}.
\newblock In {\em LDOW}, 2014.

\end{thebibliography}

\end{document}